\documentclass[10pt,letterpaper]{article}
\pdfoutput=1
\usepackage{opex3}
\usepackage[latin1,utf8]{inputenc}
\usepackage[english]{babel}
\usepackage[space]{cite}
\addto\captionsenglish{}

\begin{document}

\title{Direct optical measurement of light coupling into planar waveguide by plasmonic nanoparticles}
\author{Antti M. Pennanen* and J. Jussi Toppari}
\address{Department of Physics, Nanoscience Center, University of Jyväskylä, FIN-40014 Jyväskylä, Finland}
\email{*antti.m.pennanen@jyu.fi}

\begin{abstract}
Coupling of light into a thin layer of high refractive index material by plasmonic nanoparticles has been widely studied for application in photovoltaic devices, such as thin-film solar cells. In numerous studies this coupling has been investigated through measurement of \emph{e.g.} quantum efficiency or photocurrent enhancement. Here we present a direct optical measurement of light coupling into a waveguide by plasmonic nanoparticles.  We investigate the coupling efficiency into the guided modes within the waveguide by illuminating the surface of a sample, consisting of a glass slide coated with a high refractive index planar waveguide and plasmonic nanoparticles, while directly measuring the intensity of the light emitted out of the waveguide edge. These experiments were complemented by transmittance and reflectance measurements. We show that the light coupling is strongly affected by thin-film interference, localized surface plasmon resonances of the nanoparticles and the illumination direction (front or rear). 
\end{abstract}


\section{Introduction}

Localized surface plasmons (LSPs) on metallic nanoparticles have since the beginning of this millennium been considered for enhancement of the efficiency of photovoltaic devices such as photodetectors \cite{Stuart1996,Stuart1998,Stuart1998a,Schaadt2005,Sundararajan2008,Atwater2010}, light emitting diodes \cite{Pillai2006} and, perhaps most importantly, solar cells \cite{Schaadt2005, Catchpole2006, Derkacs2006, Pillai2007a, Lim2007, Catchpole2008, Catchpole2008c, Hagglund2008, Nakayama2008, Sundararajan2008, Akimov2009, Beck2009, Mokkapati2009, Temple2009, Atwater2010, Beck2010, Ouyang2010, Tsai2010, Beck2011, Liu2011, Mokkapati2011, Ouyang2011, Pillai2011, Paris2012, Villesen2012, Yang2012}.
Two principal mechanisms for enhancement of the efficiency of solar cells by plasmonic particles have been identified\cite{Atwater2010}: (i) improved charge carrier generation caused by the near field of LSPs on metal nanoparticles embedded in photovoltaic material, and (ii) light trapping via scattering of light by LSPs on nanoparticles located on the front or rear surface of a solar cell. The latter has been proven to enhance the efficiency of thin-film solar cells, made of silicon (Si) \cite{Catchpole2006, Derkacs2006, Pillai2007a, Beck2009, Ouyang2010, Ouyang2011, Pillai2011} as well as other photovoltaic materials, such as  gallium-arsenide (GaAs) \cite{Nakayama2008, Liu2011}, by coupling light into the guided modes of the substrate, which is also the focus of this article. 
Light trapping and coupling of light into solar cell has been widely investigated by computational studies and by measuring the enhancement of the photocurrent or the quantum efficiency caused by nanoparticles. But, to our knowledge, up to this day, direct optical measurement of light coupling into the guided modes within a waveguide has not been conducted. Here we present such measurement.

\section{Methods}

\begin{figure}
 	\includegraphics[width=1 \textwidth]{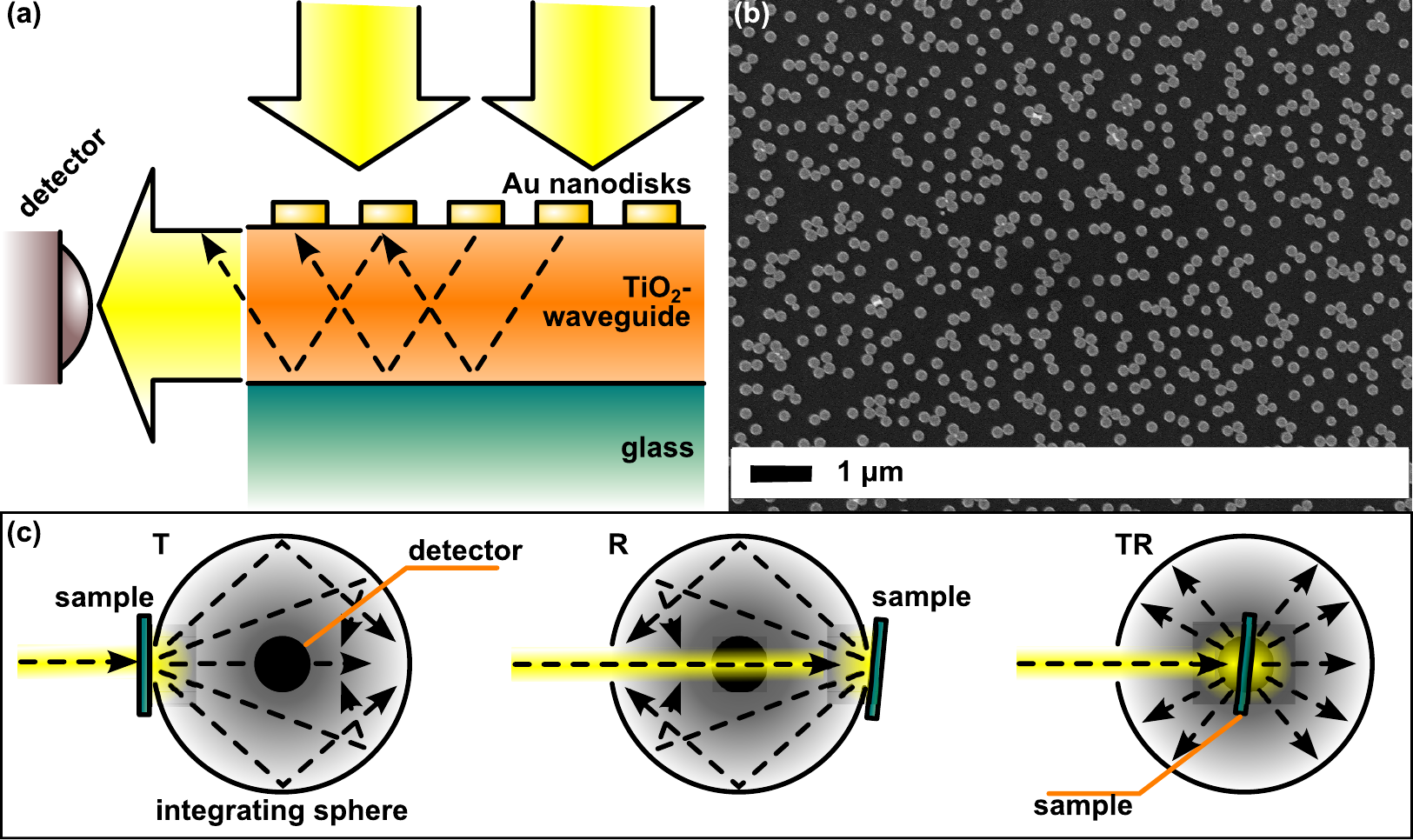}
	\caption{(a) Illustration of the optical measurement geometry in the front illumination configuration. (b) SEM image of the 42 nm $\times$ 200 nm (height $\times$ diameter) Au nanodisks on a TiO$_2$-coated glass slide. (c) Measurement of total transmittance (T), reflectance (R) and transflectance (TR) with an integrating sphere. All transmitted/reflected light entering the integrating sphere is eventually collected into the detector, situated at the bottom of the sphere.}
	\label{Fig:schematic}
\end{figure}

\subsection{Sample fabrication}

Samples consisted of gold (Au) nanodisks deposited either directly on a 1 mm thick soda-lime glass substrate or a glass substrate coated with a thin layer of titanium dioxide (TiO$_2$), as illustrated in Fig. \ref{Fig:schematic}(a). TiO$_2$ layers, acting as a planar 2D-waveguides, with thickness ranging from $100$ to $900$ nm were grown on the glass substrates by atomic layer deposition (ALD). Samples were then cut to approximately $2 \times 2$ cm$^2$ size. Nanoparticle coatings of random arrays of circular gold nanodisks with height of 42 nm and diameter of 200 nm were fabricated by hole-mask colloidal lithography (HCL) \cite{Fredriksson2007}. Surface coverage of the particles was on average (21$\pm$1)\%. A scanning electron microscope (SEM) image of the Au particles on a  TiO$_2$-coated glass slide is presented in Fig. \ref{Fig:schematic}(b). Compared to other easily fabricable shapes, such as spheres, nanodisks have high coupling efficiency to substrate \cite{Catchpole2008c, Beck2011}. Gold was  selected as preferred particle material because of its stability compared to for example silver, despite its higher absorption losses and slightly lower scattering cross-section. TiO$_2$ was used as the waveguide material because of its high refractive index and low optical absorption, to enable direct optical measurement of the waveguide coupling. 

\subsection{Measurements} 

The measurement geometry for investigation of the coupling efficiency is illustrated in Fig. \ref{Fig:schematic}(a). A broadband light source was used to illuminate the sample through a monochromator. Incident light is coupled into the waveguide by the nanoparticles and light emitted out of the waveguide edge is measured by photodiode. Measurements performed with this geometry will henceforth be referred to as emission measurements.  We use the relative intensity of the emitted light compared to the incident intensity as a measure of coupling efficiency into the guided modes of the waveguide. It should be emphasized that by coupling efficiency we mean here the fraction of all light incident to sample surface that is coupled into the guided modes of the waveguide by the particles, in contrast to some references, where coupling efficiency is defined as the fraction of light scattered by particles that is scattered into the substrate.  These emission measurements were carried out in two different configurations: the front illumination and the rear ilumination configuration. The front illumination configuration, where the sample is illuminated from the particle-coated side, is presented in Fig. \ref{Fig:schematic}(a). In the rear illumination configuration light is incident from the opposite side of the sample, \emph{i.e.} substrate side, but otherwise the geometry remains the same as in Fig. \ref{Fig:schematic}(a). 

An Oriel model 66182 light source with an USHIO halogen projector lamp was used for sample illumination, together with Acton SP2150 monochromator equipped with a 750 grooves/mm grating. Emission was measured with a Thorlabs PDA30G-EC AC-coupled amplified PbS photodiode (PD). To eliminate noise from the measurement we used an optical fork chopper in front of the monochromator entrance slit and a lock-in amplifier with the  photodetector. Furthermore, a custom made slit with gold coated blades was installed into the front collar of the PD and sample edge was squeezed between the blades to exclude any external light from measurement and to efficiently collect the light emitted out of the waveguide edge by reflection from the gold coated inner surfaces of the blades.  To eliminate higher order reflections from the illuminating light beam, long-pass filters with cut-off wavelengths of $600$ and $1200$ nm were used at the entrance of the monochromator. 

To study the extinction and plasmonic properties of the particles, measurements of beam transmittance were performed on the samples with the equipment described above. The PD was placed behind the sample, on the path of the illumination beam.

Total (beam plus diffuse) transmittance and reflectance, as well as total transflectance measurements, were carried out with a dual beam Perkin Elmer Lambda 1050 UV-Vis-NIR spectrophotometer equipped with an integrating sphere accessory. These measurements are illustrated in Fig. \ref{Fig:schematic}(c). In transmittance measurements the sample was placed at the entrance port of the integrating sphere and the light transmitted through the sample was collected by the sphere into the detector. For reflectance measurements the sample was placed behind the sphere, at the reflectance measurement port, deflected 8 degrees from normal incidence angle, and the reflected light was collected by the sphere. Transflectance measurements were carried out by inserting the sample into the sphere by using center sample mount, deflected 8 degrees from normal incidence angle, allowing both transmitted and reflected light to be collected into the detector by the integrating sphere.

\section{Results}

\subsection{Emission measurements}

\begin{figure}
	\centering
	\includegraphics[width=.5 \textwidth]{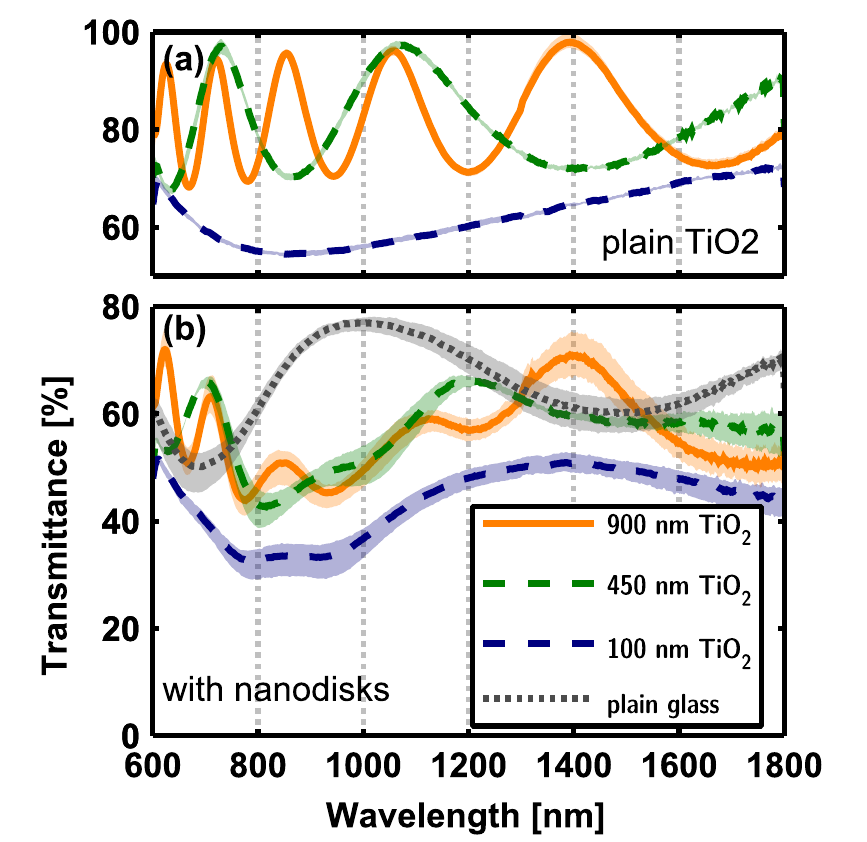}
	\caption{(a) Transmittances of TiO$_2$ films without Au nanodisks and (b) transmission spectra of samples with nanoparticles on plain glass and on 900, 450 and 100 nm TiO$_2$. Bands around lines show the maximum error limits of measurements}
	\label{Fig:transmittances}
\end{figure}

Bottom part of Fig. \ref{Fig:transmittances} shows measured transmittances of samples with Au disks on plain glass and on 100, 450 and 900 nm thick TiO$_2$ waveguides. The random particle arrays exhibit two discrete LSP resonances; the one on shorter wavelength is situated just under 700 nm for the particles on plain glass substrate and is red shifted to around 850 nm for particles on high refractive index TiO$_2$ substrate (n$\approx$2.3), and the other one on longer wavelength at around 1450 nm for particles on plain glass and above 1800 nm for particles on TiO$_2$. In the transmission spectra of samples with  TiO$_2$, the effect of thin-film interference may clearly be observed. In 450 and 900 nm films the effect is more noticeable, whereas in 100 nm film the variation as a function of wavelength is very slow and the effect, therefore, is not so prominent. Thin-film interference of plain TiO$_2$ films without nanoparticles is shown in top part of Fig. \ref{Fig:transmittances}. All transmittances shown in Fig. \ref{Fig:transmittances} are averages of three separate measurements of different samples with similar coatings (particles and/or waveguide layer). 

\begin{figure}
	\includegraphics[width=1 \textwidth]{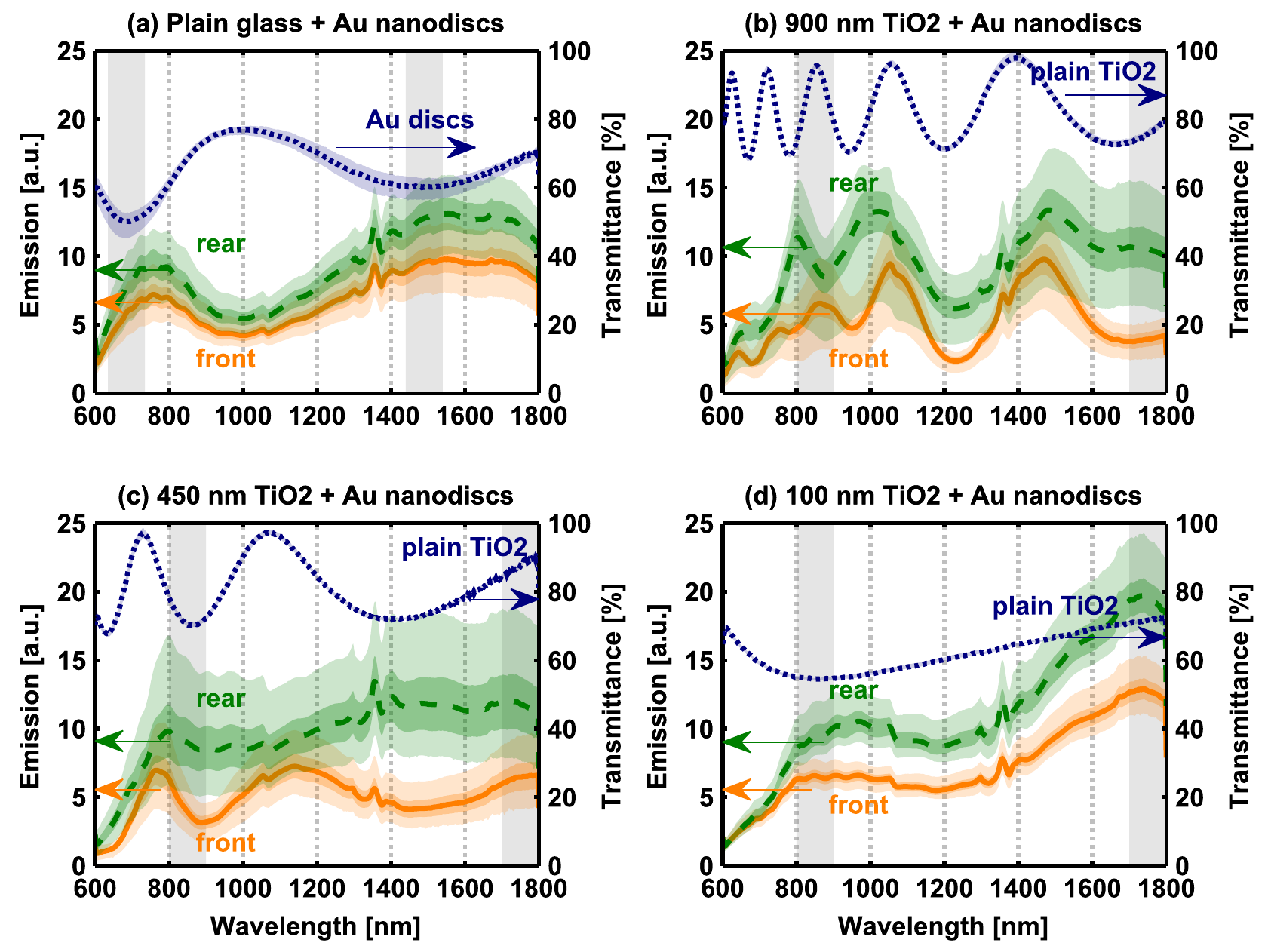}
	\caption{Emission spectra, \emph{i.e.} intensity of light emitted out of the waveguide edge in front (solid orange line) and rear (dashed green line) illumination configurations, of samples with Au nanodisks on (a) plain glass and (b) 900, (c) 450 and (d) 100 nm TiO$_2$. Dotted lines: Transmittance of the Au particles on glass is shown for reference in (a) and transmittance of the TiO$_2$ films without the nanoparticles in (b--d), to illustrate the correspondence between thin-film interference and coupling efficiency.  Bands around lines show the 95 \% confidence bands (inner bands) and maximum errors (outer bands) of measurements. If only one band is visible, it is the maximum error band. Vertical bands show locations of the resonant LSP modes.}
	\label{Fig:emissions}
\end{figure}

In the emission measurements, the intensity of the light emitted out of the waveguide edge was observed to be very low; less than 1\% of the intensity of the beam illuminating the sample surface. This is attributed to cumulated absorption in the nanoparticles, which causes attenuation of the light propagating in the waveguide. Figures \ref{Fig:emissions}(a-d) illustrate the coupling of light into waveguides by the plasmonic particles, \emph{i.e.} the intensity of light emitted out of the sample edge, in the direct optical measurement arrangement shown in Fig. \ref{Fig:schematic}. All emission spectra are averages of total of 9 to 12 separate measurements of three different samples with similar coatings. All emissions are normalized with the spectrum of the light source, which was measured with the same monochromator and detector as the emission from the sample edge. Figure \ref{Fig:emissions}(a) has the Au nanodisks directly on top of a plain glass substrate and Fig. \ref{Fig:emissions}(b-c) on top of 900, 450 or 100 nm thick TiO$_2$ waveguide layer, respectively. Emissions under front and rear illumination are shown. 

The 1 mm thick glass substrate also itself acts as a waveguide for the light scattered by the Au particles, as may be observed from Fig. \ref{Fig:emissions}(a). The glass slide acts as a waveguide also when an additional high refractive index waveguide layer is present. Particles directly on a glass substrate exhibit largest coupling efficiencies near the surface plasmon resonance wavelengths, as do the samples  with 100 nm TiO$_2$, for both front and rear illumination configurations. 

For samples with 900 and 450 nm thick waveguides, however, the effect of thin-film interference begins to add complexity to the emission spectra. For the front illumination configuration the coupling is most efficient at the wavelengths corresponding to a constructive thin-film interference in the transmittance spectra. Emission spectra of samples with 900 nm TiO$_2$ films show some interference-like pattern in the rear illumination configuration also, but the number and positions of the emission peaks are different to the measurement with front illumination. Samples with 450 nm waveguides do not show interference pattern in the emission spectrum under rear illumination.

\begin{figure}
	\centering
	\includegraphics[width=0.5 \textwidth]{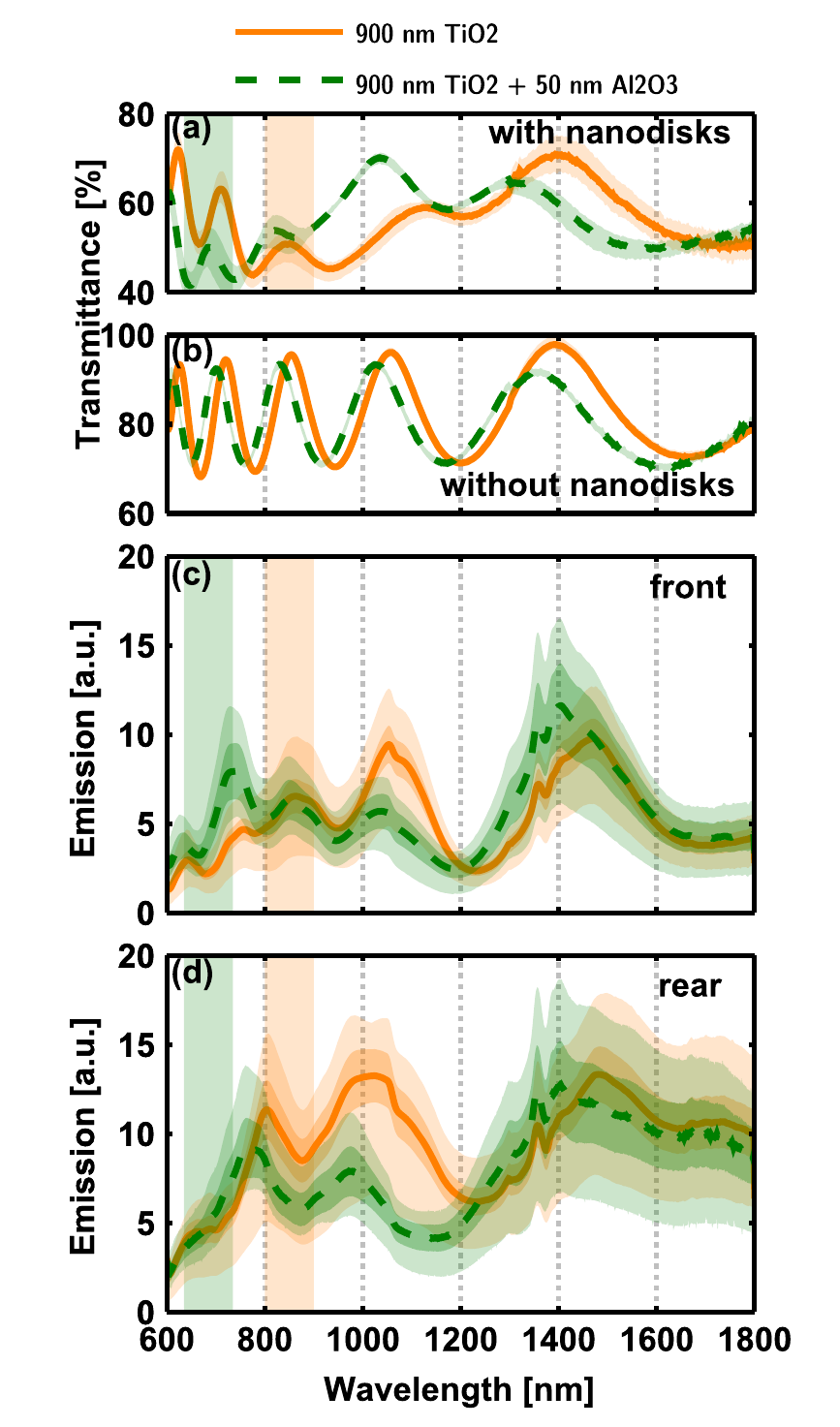}
	\caption{Effect of spacer layer on the coupling efficiency. Samples with 900 nm TiO$_2$ waveguide and Au nanodisks with (dashed green lines) and without (solid orange lines) the spacer layer. (a) Transmittances. (b) Transmittances without the nanodisks. (c,d) Emission spectra under front and rear illumination, respectively. Bands around lines show the 95 \% confidence bands (inner bands) and maximum errors (outer bands) of measurements. If only one band is visible, it is the maximum error band. Vertical bands show the locations of the shorter wavelength LSP resonances.}
	\label{Fig:K102}
\end{figure}

Effect of a spacer layer on the coupling efficiency was investigated by adding an ALD deposited 50 nm thick aluminum oxide (Al$_2$O$_3$) layer between the particles and the 900 nm TiO$_2$ waveguide (Fig. \ref{Fig:K102}). Refractive index of the Al$_2$O$_3$ film is n$\approx$1.6. The spacer layer of 50 nm is thick enough to substantially shift the localized surface plasmon resonances of the Au nanodisks (Fig. \ref{Fig:K102}(a)), but causes significantly smaller shift in thin-film interference maxima and minima (Fig. \ref{Fig:K102}(b)). 

With front illumination the coupling efficiency into the guided modes of the waveguide is increased or decreased on different wavelengths when spacer layer is added, but there is no noticeable overall increase or decrease. When illuminated from rear, there is a net decrease in coupling efficiency, with addition of the spacer layer. These results are in agreement with those presented in previous studies \cite{Beck2010, Ouyang2010, Beck2011, Pillai2011, Yang2012}.

\subsection{Transmittance and reflectance measurements} 

\begin{figure}
	\centering
	\includegraphics[width=0.5 \textwidth]{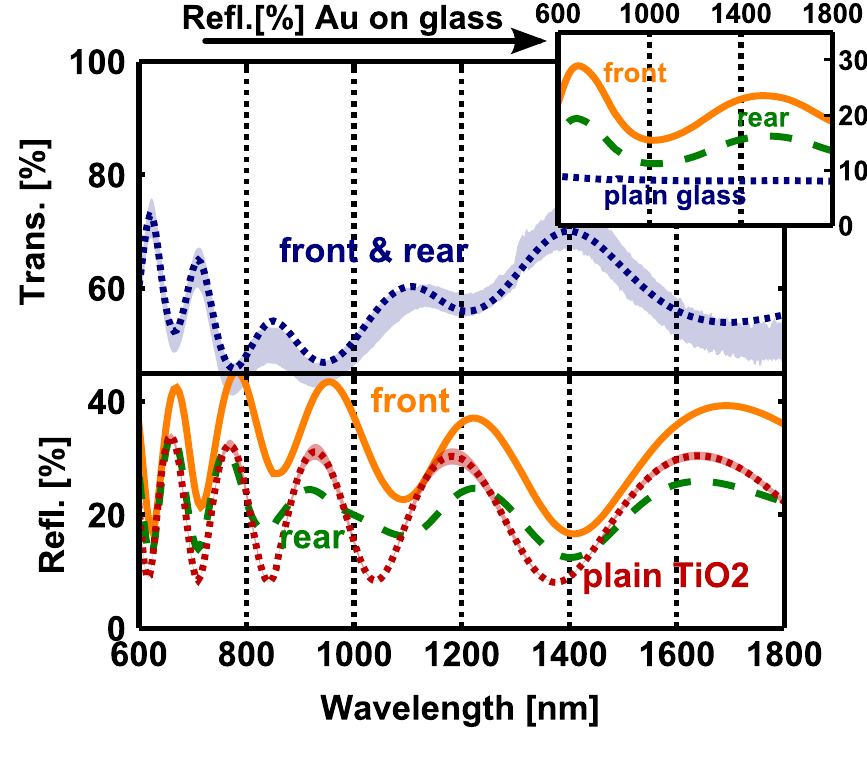}
	\caption{Sample with $900$ nm TiO$_2$ and Au nanodisks. Upper part: total transmittance (dotted blue line) and beam transmittance with maximum error limits (band around dotted line).  Lower part: Total reflectance for front (solid orange line) and rear illumination (dashed green line) and for sample without Au nanodisks (dotted red line). Inset: total reflectance of Au nanodisks on plain glass.}
	\label{Fig:trans+refl}
\end{figure}

In order to gain additional insight into the asymmetry in coupling efficiency between front and rear illumination, measurements with a dual-beam spectrophotometer and an integrating sphere were carried out. One individual sample with $900$ nm TiO$_2$ waveguide and 42 $\times$ 200 nm circular Au disks was selected for these measurements. Results of the total transmittance and reflectance measurements, with both front and rear illumination, are shown in Fig. \ref{Fig:trans+refl}. Inset shows measured reflectances for Au nanodisks on plain glass for reference. 

On plain glass, the Au nanodisks increase the reflectance, and reflectance is significantly larger for front than rear illumination. Au particles on TiO$_2$ coated glass increase the reflectance when the sample is illuminated from the front and cause the reflectance to increase or decrease, depending on the wavelength, when illuminated from the rear. In this respect our samples behave differently compared to Si solar cells, where nanoparticles act as an anti-reflection coating, when deposited on the front surface of the cell \cite{Pillai2007a, Beck2010, Pillai2011}. This difference is due to high transparency of the glass/TiO$_2$ substrate which is  always decreased by addition of nanoparticles (for more deatailed discussion, see section \ref{asymdiscussion}). On TiO$_2$ coated glass, the transmittance is essentially the same for both front and rear illumination. Beam transmittance of three individual samples is shown as a band around the total transmittance plot (from the one selected sample). In this case the total transmittance is essentially equal to the beam transmittance,\emph{ i.e.} the diffuse transmittance is negligible. The measured total reflectance of the plain TiO$_2$ film is shown for reference. Difference between front and rear reflectance of the plain TiO$_2$ sample is negligible. 

\begin{figure}
	\centering
	\includegraphics[width=0.5 \textwidth]{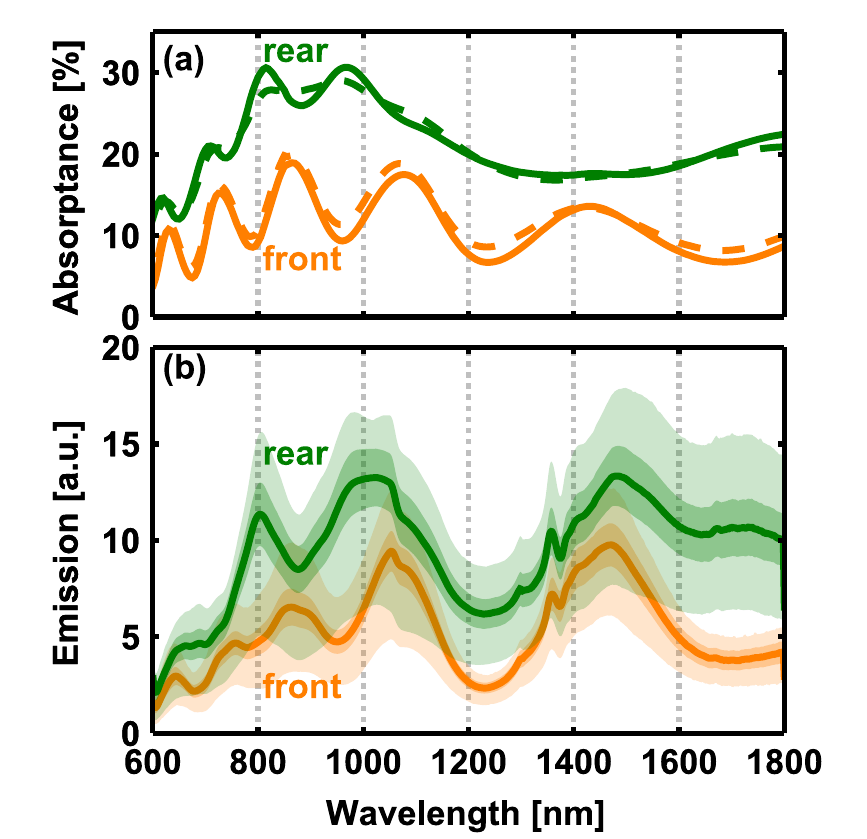}
	\caption{(a) Absorption of sample with $900$ nm TiO$_2$ and Au nanodisks. Solid lines: absorptance from the separate total transmittance and reflectance measurements; dashed lines: absorptance from the transflectance measurements. (b)  Emission spectrum of same sample, same as in Fig. \ref{Fig:emissions}(b).}
	\label{Fig:ems+abs}
\end{figure}

Mechanisms behind the electro-magnetic coupling between the nanoparticles and waveguide modes are further investigated in Fig. \ref{Fig:ems+abs}. Absorption in the samples was calculated from the total reflectance and transmittance measurements as (Abs) $=$ 100 \%$-$((Trans)+(Refl)) and from the total transflectance measurements as (Abs) $=$ 100 \%$-$(Transfl). Absorption spectra in Fig. \ref{Fig:ems+abs}(a) show apparent similarity in shape and relative magnitude to the emission spectra in Fig. \ref{Fig:ems+abs}(b). 

\section{Discussion}

\subsection{Effect of plasmon resonance and thin-film interference}

Au particles on plain glass substrate and on the thinnest 100 nm TiO$_2$ waveguide show largest coupling efficiency near the wavelengths of the localized surface plasmon resonances, with both front and rear illumination. The effect of the very thin TiO$_2$ waveguide is to red-shift the plasmon resonance (as can be observed from Fig. \ref{Fig:transmittances}), and thus the peaks in coupling efficiency, and to increase the coupling efficiency near the second, longer wavelength plasmon resonance. Looking at the transmittance spectrum of plain 100 nm TiO$_2$ film, this is probably due to constructive thin-film interference near the longer localized surface plasmon resonance wavelength. What is notable here is that the 100 nm TiO$_2$ waveguide does not support guided modes at wavelengths above around 850 nm. Therefore, above this wavelength, all emission comes from the glass substrate acting as a waveguide. Emission enhancement from plain glass to 100 nm TiO$_2$ coated glass may be (at least partly) attributed to the change in dielectric environment of the particle: first, the particle scattering tends to be directed more strongly to the larger refractive index, and second, the scattering efficiency of the particle may be increased by the higher refractive index environment \cite{Catchpole2008}. The presence of the very thin TiO$_2$ film may also direct the scattering into directions more preferable for the guided modes in the glass substrate, \emph{i.e.} to angles more parallel to the waveguide.

With 900 and 450 nm TiO$_2$ films also the thin-film interference significantly affects the coupling efficiency. Based on the behaviour of the emission in samples with no or 100 nm TiO$_2$, it seems most plausible that the coupling is defined by the plasmon resonances but modified by the driving field, which is defined by the thin-film interference. It should be noted, that the thin-film interference effect does not manifest itself in heavily absorbing waveguides, where the optical path length is shorter than the device thickness. But even in Si or GaAs solar cells interference effects become significant when the cell thickness is low and at wavelengths where the optical absorption is weak, so that optical path lengths of several times the device thickness are required for efficient photocurrent generation at the corresponding wavelength. What is notable here is that the coupling efficiency of incident light into the guided modes of the substrate is most prominently dependent on the thin-film interference and not, for example, the waveguide modes, contrary to what has been observed in some previous studies \cite{Stuart1998, Pillai2006}.

\subsection{Effects of near and far fields}

In the emission spectra of Fig. \ref{Fig:emissions}, the short and long wavelength localized surface plasmon modes both contribute to the coupling of light into the waveguide. Some numerical studies\cite{Hagglund2008,Beck2011} suggest that for disk shaped metal nanoparticles on a high refractive index substrate, the discrete resonances at different wavelengths are associated with different field distributions. Beck \emph{et al.}\cite{Beck2011} observe two separate resonances in Ag disks; on shorter wavelengths the resonance is associated with modes localized at the particle-air interface, while the longer wavelength resonances involve modes at the particle-substrate interface.  Hägglund \emph{et al.}\cite{Hagglund2008} argue that the coupling efficiency to waveguide is determined by the far field  whereas Catchpole \emph{et al.}\cite{Catchpole2008c} and Beck \emph{et al.}\cite{ Beck2011} emphasize the importance of near-field coupling. We observe no significant difference in strength of the coupling to the waveguide between the shorter and longer wavelength plasmonic modes, even though according to Hägglund \emph{et al.} \cite{Hagglund2008} as well as Catchpole \emph{at al.} \cite{Catchpole2008c} and Beck \emph{at al.}\cite{ Beck2011} the near field localizations are completely different for the two resonances. Thus it seems that, as Hägglund \emph{at al.}\cite{Hagglund2008} suggested, the coupling into the waveguide is not affected by the different near field distributions associated with the individual plasmonic modes, but is determined by the far field.

Particles deposited on the front surface of a photovoltaic cell may enhance or suppress the photocurrent and efficiency of the device, depending on the wavelength.  Suppression of the photocurrent and the decrease of efficiency of Si solar cells  at just below LSP resonance wavelength, due to the destructive interference between incident and scattered fields, has been reported in many studies \cite{Lim2007, Sundararajan2008, Beck2009, Beck2010, Liu2011}. Furthermore, constructive or destructive interference may arise from different field distributions associated with the different resonant modes, causing either positive or negative effect on the solar cell efficiency \cite{Hagglund2008}. In our measurements, the emission spectra suggest that coupling occurs also for wavelengths just below the first, shorter wavelength resonance, as well as the second, longer wavelength resonance. Effect of this coupling would be to increase the optical path length of photons and thus enhance photocurrent in photovoltaic material. However, the field scattered by the particles may be at opposite phase with the incident field, as suggested in numerous studies\cite{Lim2007, Sundararajan2008, Beck2009, Beck2010, Liu2011}, causing decrease in field intensity near the particle. The net effect on the photocurrent generation may thus be negative, as observed in references mentioned above.

\subsection{Asymmetry between front and rear illumination}
\label{asymdiscussion}

Negative effects of destructive interference on solar cell efficiency may  be avoided by placing the particles on the rear surface of the cell \cite{Beck2009, Beck2010, Ouyang2010, Ouyang2011, Pillai2011}. Placement of the particles to the rear of the cell also avoids the negative effect of particle absorption to the cell performance, compared to front deposition \cite{Catchpole2006, Ouyang2011}. Particles located on the rear surface of a thin-film Si cell have been reported to provide larger quantum efficiency enhancement than front located particles \cite{Beck2009, Ouyang2011}. In some references it is reported that there is no significant difference in the fraction of light scattered by the particles that is scattered into the substrate between front and rear located particles \cite{Beck2010, Pillai2011}, while there is a notable asymmetry in another reference \cite{Mokkapati2011}. These differences may be due to different particle sizes simulated in different references. In the measurements presented here the coupling efficiency to the guided modes within the waveguide, measured as the emission from the sample edge, is higher for rear illumination for all samples (see Fig. \ref{Fig:emissions}). Thus, it seems that for Au nanodisks used here, the fraction of incident light that is coupled into the guided modes of the substrate does depend on the illumination direction. The difference in coupling efficiency to guided modes between front and rear illuminations observed in our measurements is larger for the samples with  TiO$_2$, compared to that with just plain glass. This would suggest that the higher refractive index environment of the nanoparticles causes the asymmetry in coupling efficiency between front and rear illumination configurations to be increased. This interpretation is further supported by the measurements of the samples with a 50 nm Al$_2$O$_3$ spacer layer; in Fig. \ref{Fig:K102} asymmetry in the coupling efficiency is decreased when the lower refractive index Al$_2$O$_3$ spacer layer is introduced between the high index TiO$_2$ substrate and the nanoparticles.

To further investigate the asymmetry in the coupling efficiency, measurements of reflectance, transmittance and transflectance were conducted with an integrating sphere on one selected sample with 900 nm TiO$_2$ and Au nanodisks. In Fig. \ref{Fig:trans+refl} total transmittances are equal in front and rear illumination. Total transmittance is also essentially equal to the beam transmittance, indicating that all transmittance is beam. This would suggest that the scattering into the forward direction by the particles is zero. Total reflectances, however, are different for front and rear illumination, implying that backward scattering is affected by the direction of illumination. This may be due to the asymmetry in refractive index around the particle, as the scattering tends to be directed towards the higher index. When illuminated from the rear, particles tend to scatter more efficiently into the backward direction, but this backscattered light is then coupled efficiently into the waveguide, resulting in the decrease of reflectance. This interpretation, however, does not explain the absence of forward scattering in the transmittance measurements. 

\begin{figure}
	\centering
	\includegraphics[width=0.5 \textwidth]{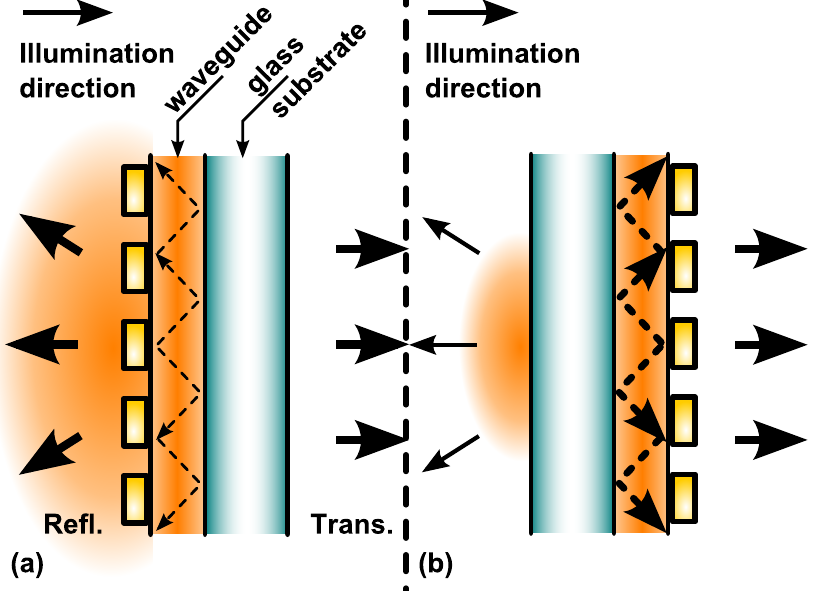}
\vspace{-2mm}
	\caption{Transmission and reflection illustrated with (a) front and (b) rear illumination. Reflectance is decreased when the sample is illuminated from the rear, whereas transmittance remains unchanged.}
	\label{Fig:scattering}
\end{figure}

In the emission measurements, the intensity of light emitted out of the waveguide edge was found to be very low. We attributed this to the cumulated absorption in the nanoparticles. This interpretation is supported by Fig. \ref{Fig:ems+abs}, where absorptance of the sample with 900 nm TiO$_2$ coating is 30 \% at its highest for rear and 20 \% for front illumination. The absorptance of plain 900 nm TiO$_2$ film is negligible. Furthermore, the features and relative magnitude of absorptances under front and rear illumination in Fig. \ref{Fig:ems+abs}(a) are similar to those of corresponding emission spectra in Fig. \ref{Fig:ems+abs}(b).  This may be understood as follows: When light, trapped into the waveguide, propagates and interacts with the nanoparticles, an individual particle may interact with it by either absorption or scattering. Since the refractive index of the TiO$_2$ substrate is much higher than that of air, scattering is predominantly directed into the waveguide. But even if the particles scatter with much higher efficiency than they absorb, the light propagating in the waveguide interacts with very high number of particles, leading to significant absorption loss. There is thus a correspondence between the coupling efficiency and the sample absorptance. Since the transmittance is independent of the illumination direction, the asymmetry in absorptance and coupling efficiency between front and rear illumination is related to the asymmetry in reflectance,\emph{ i.e.} back scattering. This interpretation is illustrated in Fig. \ref{Fig:scattering}: transmission through the sample remains the same, regardless of the illumination direction, but the reflectance and the fraction of incident light coupled into the waveguide changes. What is notable here is that the front reflectance is higher than the rear reflectance: on Si cells, front located nanoparticles typically act as an antireflection coating, reducing the reflectance,  but the antireflection property is absent for rear located particles \cite{Beck2010}. In our measurements, the reflectance of the sample with Au disks
is actually higher than that of plain TiO$_2$,  as seen in Fig. \ref{Fig:trans+refl}. This is explained by the fact that in our samples the substrate and the waveguide are highly transparent, contrary to the Si solar cells, and thus the introduction of metal nanoparticles onto the sample surface decreases the transmittance. Under front illumination, this is observed as increased reflectance at all wavelengths. Together with decreased transmittance, the effect of the Au nanodisks is to couple incident light into the waveguide, leading to absorption losses. As observed in Fig. \ref{Fig:emissions}, this coupling is more efficient when illuminating the sample from rear, which explains the decrease in reflectance at some wavelengths and increase at others under rear illumination. 

\subsection{Effect of spacer layer}

Addition of a 50 nm Al$_2$O$_3$ spacer layer between the particles and the high index waveguide causes a substantial
blueshift to the resonances, but also a much smaller shift to the interference pattern in the transmittance of the bare TiO$_2$ film. 
The shift in the resonance peaks results in a shift of the enhancement peaks in the emission spectra for both front and rear illumination. The small shift in the thin-film interference pattern causes a small shift to the interference pattern of the emission spectra also. The most prominent effect of the addition of the spacer layer on the coupling efficiency in the front illumination configuration is to shift the position of the enhancement peaks, due to the blueshift of the plasmonic resonance. With rear illumination, however, the addition of the spacer layer results in noticeable decrease of emission at wavelengths of 800--1200 nm. Our measurements are in agreement with previous results\cite{Beck2010, Ouyang2011, Pillai2011} that rear located particles exhibit highest coupling efficiency when deposited directly on a high index substrate.

\section{Conclusions}

Here we present the first direct optical measurement of the coupling of light into a planar waveguide by plasmonic nanoparticles. As a conclusion, we have observed that if the high-index waveguide is situated on a transparent substrate or superstrate, the thin-film interference significantly affects the coupling between the particle plasmons and waveguide modes. This arises from the differences in strength of the electromagnetic field driving the particle plasmons, caused by the interference effects. These effects could be significant in thin-film photovoltaic cells at wavelengths near the band gap, where optical absorption is weak. It seems that the efficiency of light coupling into the guided modes is most importantly determined by the thin-film interference and the plasmon resonances, and not, for example, the waveguide modes, contrary to what has been reported in some previous studies.\cite{Stuart1998, Pillai2006} Au nanodisks used in this study exhibited two distinct localized plasmon resonances, both of which demonstrated coupling into the waveguide.

We have also noted that the coupling efficiency into the guided modes is different under front and rear illuminations, which is in agreement with results of other studies \cite{Beck2009, Beck2010, Mokkapati2011, Ouyang2011, Pillai2011}. With no spacer layer, the coupling efficiency is greater for rear illumination, but addition of a 50 nm Al$_2$O$_3$ spacer layer decreases the efficiency in rear illumination configuration. The difference in the coupling efficiency between front and rear illumination is closely associated with the difference in the total reflectance of the nanoparticle-waveguide system.

\section*{Appendix A: Au nanodisk fabrication}

In the HCL\cite{Fredriksson2007} process, the glass slides were first spin coated with 495 PMMA-A2 for 1 minute at 2000 rpm and baked for 10 mins at 170\textdegree C.  The PMMA layer was then made hydrophilic by oxygen plasma treatment in reactive ion etch (RIE), with 50 W power for 5 seconds. Samples were then covered with 0.2 weight\% PDDA (poly-diallyldimethylammonium chloride) solution for 30 secs, rinsed with de-ionized water for another 30 secs and dried by pressing gently with a clean room sheet; then covered for 2 mins with 0.2 wt\% water solution of polystyrene (PS) nanospheres with 200 nm diameter, rinsed with DI water for 1 min and dried with a clean room sheet. Clean room sheets were used for drying instead of blowing with N$_2$ to prevent sedimentation of the PS particles on the relatively large sample surface. A mask layer was produced by first evaporating approximately 15 nm of Au under ultra high vacuum and then stripping off the PS nanospheres by transparent tape. The PMMA layer was etched through the mask by O$_2$ plasma in RIE for 1 min 12 secs with 50 W power. The nanoparticles were deposited by evaporating 42 nm of Au under ultra high vacuum. Lift-off procedure of the sacrificial PMMA layer was performed by ultrasonicating the samples for 5 mins in boiling acetone. Finally, the samples were rinsed with isopropanol and DI water and blown dry with N$_2$.

\section*{Acknowledgments}

We would like to thank the Foundation for Research of Natural Resources in Finland (Suo\-men Luon\-non\-va\-rain Tut\-ki\-mus\-sää\-tiö) for funding; J. Mau\-la, Beneq for the ALD depositions; H. Tuo\-vi\-nen, University of Eastern Finland, Joen\-suu for arranging and J. Hil\-tu\-nen, SIB-labs for conducting measurements with spectrophotometer and integrating sphere.

\end{document}